\documentclass{PoS}

\usepackage{amsmath,amssymb}
\def\Tr{\mbox{Tr}}
\def\1{1\!\!\!\!\!\bot}
\usepackage{tikz}
\usetikzlibrary{plotmarks}

\title{Ghost Sector in Minimal Linear Covariant Gauge}

\ShortTitle{Ghost Sector in Minimal LCG}

\author{\speaker{Attilio~Cucchieri},$^a$
        David~Dudal,$^{bc}$
        Tereza~Mendes,$^a$
        Orlando~Oliveira,$^d$ \vskip 1mm
        Martin~Roelfs$^{b}$
        and Paulo~J.~Silva$^d$ 
        \vskip 1.5mm \\
        \llap{$^a$} Instituto de F\'\i sica de S\~ao Carlos, Universidade de S\~ao Paulo,
        C.P.\ 369, 13560-970 S\~ao Carlos, SP, Brazil
        \vskip 1mm \\
        \llap{$^b$} KU Leuven Kulak, Department of Physics,
        Etienne Sabbelaan 53 bus 7657, 8500 Kortrijk, Belgium
        \vskip 1mm \\
        \llap{$^c$} Ghent University, Department of Physics and Astronomy,
        Krijgslaan 281-S9, 9000 Gent, Belgium
        \vskip 1mm \\
        \llap{$^d$} CFisUC, Departamento de F\'\i sica, Universidade de Coimbra,
        3004-516 Coimbra, Portugal
        \vskip 1mm \\
        E-mail: \email{attilio@ifsc.usp.br}, \email{david.dudal@kuleuven.be},
        \email{mendes@ifsc.usp.br}, \email{orlando@fis.uc.pt},
        \email{martin.roelfs@kuleuven.be}, \email{psilva@uc.pt}}

\abstract{
We discuss possible definitions of the Faddeev-Popov matrix
for the minimal linear covariant gauge on the lattice and
present preliminary results for the ghost propagator.
}

\FullConference{XIII Quark Confinement and the Hadron Spectrum - Confinement2018\\
		31 July - 6 August 2018\\
		Maynooth University, Ireland}

\begin{document}

%%%%%%%%%%%%%%%%%%%%%%%%%%%%%%%%%%%%%%%%%%%%%%%%%%%%%%%%%%%%%%%%%%%%%%%%%%%%%%%%%%%%%

\section{Why Study the Linear Covariant Gauge on the Lattice?}

A big effort has been made in the last decades in order
to understand the infrared properties of Yang-Mills theories
from the study of their Green's functions, notably the
gluon and the ghost propagators (see, e.g.,
\cite{Greensite:2011zz} and references therein).
To this end, several gauge-fixing conditions have been
considered: Landau gauge, Coulomb gauge, $\lambda$-gauge,
maximal Abelian gauge, etc.
In particular, in Landau gauge, the so-called Gribov-Zwanziger
(GZ) confinement scenario has captivated a lot of attention
\cite{Vandersickel:2012tz}.
A natural generalization of Landau gauge is the linear
covariant gauge (LCG).
Recently, a possible way of extending the GZ approach (in the
continuum) to LCG has been proposed \cite{Capri:2015pja}.
The resulting theory is characterized by an exact nilpotent
nonperturbative BRST symmetry \cite{Capri:2015ixa}.

For the moment, numerical simulations have focused
on the implementation of the LCG on the lattice
\cite{Giusti:1996kf,Giusti:1999im,LCGlattice,Cucchieri:2011aa}
and on the study of the gluon propagator \cite{LCGlattice,
LCGgluon2,LCGgluon3}.
On the other hand, it is still an open problem how to
define the Faddeev-Popov (FP) matrix for LCG on the lattice.
This is the goal of our study.
Preliminary results have been presented in \cite{Cucchieri:2018doy,
Cucchieri:2018leo}.
Some early results were also presented in \cite{thesis}.

%%%%%%%%%%%%%%%%%%%%%%%%%%%%%%%%%%%%%%%%%%%%%%%%%%%%%%%%%%%%%%%%%%%%%%%%%%%%%%%%%%%%%

\section{Minimal Linear Covariant Gauge}

In order to impose the LCG on the lattice, one should first
recall that the lattice Landau gauge, which is a special case
of the LCG, is obtained by minimizing the functional (see,
e.g., \cite{Giusti:2001xf})
\begin{equation}
\mathcal{E}_{LG}[U_{\mu},g] \,\equiv\,
- \, \Re \,\Tr \sum_{x} \sum_{\mu=1}^d \, U^{g}_{\mu}(x)
\,\equiv\, - \, \Re \,\Tr \sum_{x} \sum_{\mu=1}^d \,
g(x)\, U_{\mu}(x)\, g^{\dagger}(x+e_{\mu}) \; .
\end{equation}
One can interpret the Landau-gauge functional as a spin-glass
Hamiltonian for the ``spin variables'' $g(x)$ with a ``random
interaction'' given by the link variables $U_{\mu}(x)$.
This suggests the existence of several local minima for
$\mathcal{E}_{LG}[U_{\mu},g]$. 
The set of these local minima, known as Gribov copies
\cite{Vandersickel:2012tz,Gribov:1977wm,Zwanziger:1993dh},
defines the first Gribov region $\Omega$.
From the second variation of the minimizing functional, we
define the Landau FP operator \cite{Zwanziger:1993dh}
\begin{eqnarray}
& & \!\!\!\!\!\!\!\!\!\!
  \mathcal{M}^{bc}(x,y) \, \equiv \,
     \sum_{\mu=1}^d \, \Biggl\{ \, \Gamma^{bc}_{\mu}(x) \,
         \Bigl[ \, \delta_{x,\,y} \, - \,
   \delta_{x+e_{\mu},\,y} \, \Bigr]
 \, + \, \Gamma^{bc}_{\mu}(x-e_{\mu}) \,
        \Bigl[ \, \delta_{x,\,y} \, - \,
   \delta_{x-e_{\mu},\,y} \, \Bigr]
 \nonumber \\
& &\qquad\qquad\quad\;\; - \, \sum_{e=1}^{N_c^2 - 1} \,
     f^{bec} \Bigl[ \, A^e_{\mu}(x-e_{\mu}/2) \,
       \delta_{x-e_{\mu},\,y} \, - \,
           A^e_{\mu}(x+e_{\mu}/2)
      \, \delta_{x+e_{\mu},\,y} \, \Bigr]
                    \, \Biggr\} \; ,
\label{eq:MFP}
\end{eqnarray}
which is symmetric (under the simultaneous exchanges $b \leftrightarrow
c$ and $x \leftrightarrow y$) and semi-positive definite.
In the above equation we used the definition
\begin{equation}
\Gamma^{bc}_{\mu}(\vec{x}) \, \equiv \, \Tr \, \left[ \,
   \frac{\lambda^b \, \lambda^c \, + \, \lambda^c \,
          \lambda^b}{4} \, \frac{U_{\mu}(\vec{x}) \, + \,
          U^{\dagger}_{\mu}(\vec{x})}{2} \,
    \right] \; ,
\label{eq:defGamma}
\end{equation}
where $\{ U_{\mu}(\vec{x}) \}$ is the gauge-fixed configuration
and the matrices $\lambda^b$ are the $N_c^2 - 1$ traceless
Hermitian generators of the gauge group SU($N_c$), normalized
such that $ \Tr ( \lambda^b \lambda^c ) = 2 \delta^{bc}$.
Note that for the usual generators $\tilde{\lambda}_b$
with normalization $\Tr\, ( \tilde{\lambda}_b \tilde{\lambda}_c ) =
\delta_{bc}/2$, we have
\begin{equation}
\tilde{\lambda}_b \, =\, \frac{\lambda_b}{2} \; .
\label{eq:lambda}
\end{equation}
We also defined the lattice gauge field through the relations
\begin{eqnarray}
A_{\mu}(\vec{x}+\vec{e}_{\mu}/2) &\!\equiv\!& \frac{1}{2 \,i}\,
   \left[ \, U_{\mu}(\vec{x})
     - U_{\mu}(\vec{x})^{\dagger} \, \right] \, - \,  \1 \,
        \frac{\Tr}{2 \,i\,N_c}\, \left[ \, U_{\mu}(\vec{x})
           - U_{\mu}(\vec{x})^{\dagger} \, \right] \; ,
\phantom{ooo}
\label{eq:defAgen} \\[2mm]
A_{\mu}^{b}(\vec{x}+\vec{e}_{\mu}/2) &\!=\!&
      \frac{1}{2} \, \Tr \; [\, A_{\mu}(\vec{x}+\vec{e}_{\mu}/2)
          \, \lambda_b \, ] \label{eq:A} \; ,
\end{eqnarray}
where $\1$ is the identity matrix.

It is important to note that one can write the Landau FP operator
(\ref{eq:MFP}) as \cite{Cucchieri:2018doy}
\begin{equation}
\mathcal{M} \,=\,
\frac{1}{2}\,\left(\,
\mathcal{M}_{+} \, + \, \mathcal{M}_{-} \, \right) \; ,
\label{eq:MMM}
\end{equation}
with
\begin{equation}
\left( \mathcal{M}_{+} \, \gamma \, \right)^{b}(\vec{x})
\,\equiv\,- \, \sum_{\mu=1}^{d}\,
   \left[ \, \left( D_{\mu} \, \gamma \right)^b(\vec{x})
  - \left( D_{\mu} \, \gamma \right)^b(\vec{x}-\vec{e}_{\mu})
      \, \right]\,\equiv\, - \, \sum_{\mu=1}^{d} \, \left[ \,
 \nabla_{\mu}^{(-)} \left( \, D_{\mu} \, \gamma \, \right)
      \, \right]^b(\vec{x})
\label{eq:M+}
\end{equation}
and
\begin{equation}
\left( \mathcal{M}_{-} \, \gamma \, \right)^{b}(\vec{x})
\,\equiv\, \sum_{\mu=1}^{d} \, \left\{ \,
      D_{\mu}^{T} \, \left[ \gamma^{\,\,b}(\vec{x}+\vec{e}_{\mu})
\,-\,\gamma^{\,\,b}(\vec{x}) \right]\,\right\} \,\equiv\,
 \sum_{\mu=1}^{d} \, \left[ \,
      D_{\mu}^{T} \, \left( \nabla_{\mu}^{(+)} \, \gamma \,
           \right)\, \right]^b(\vec{x}) \; .
\label{eq:M-D}
\end{equation}
Here,
\begin{equation}
D_{\mu}^{bc}(\vec{x},\vec{y}) \,\equiv\,
  \Gamma^{bc}_{\mu}(\vec{x}) \,
  \Bigl[ \, \delta_{\vec{x}+\vec{e}_{\mu},\vec{y}}
  \, - \, \delta_{\vec{x},\vec{y}} \, \Bigr] \,
- \, \sum_{e=1}^{N_c^2-1} f^{bec} \,
   A^e_{\mu}(\vec{x}+\vec{e}_{\mu}/2) \,
\Bigl[ \, \delta_{\vec{x}+\vec{e}_{\mu},\vec{y}}
  \, + \, \delta_{\vec{x},\vec{y}} \, \Bigr]
\mbox{\phantom{ooo}}
\label{eq:Dbcxy}
\end{equation}
is the lattice gauge-covariant derivative \cite{Zwanziger:1993dh},
which implies the transpose lattice gauge-covariant derivative
\begin{eqnarray}
\left(D_{\mu}^T\right)^{bc}(\vec{x},\vec{y}) \,&\equiv&\,
\Gamma^{bc}_{\mu}(\vec{x}-\vec{e}_{\mu})\,
\delta_{\vec{x}-\vec{e}_{\mu},\vec{y}} \, - \,
\Gamma^{bc}_{\mu}(\vec{x}) \,
   \delta_{\vec{x},\vec{y}} \nonumber \\[2mm]
&&\qquad \quad \, + \, \sum_{e=1}^{N_c^2-1} \,
   f^{bec} \,\Bigl[ \,A^e_{\mu}(\vec{x}+\vec{e}_{\mu}/2) \,
   \delta_{\vec{x},\vec{y}} \, + \,
     A^e_{\mu}(\vec{x}-\vec{e}_{\mu}/2) \,
     \delta_{\vec{x}-\vec{e}_{\mu},\vec{y}} \, \Bigr] \; .
\end{eqnarray}
We have also indicated with $\nabla_{\mu}^{(+)}$ [respectively
$\nabla_{\mu}^{(-)}$] the usual forward (respectively backward)
lattice derivative.
Since the transpose of the backward lattice derivative
$\nabla_{\mu}^{(-)}$ is given by $- \nabla_{\mu}^{(+)}$, it is
evident that $\mathcal{M}_{-}^{T} = \mathcal{M}_{+}$
and the matrix $\mathcal{M}$ in Eq.\ (\ref{eq:MMM}) can
be written as $\,\left(\mathcal{M}_{+} + \mathcal{M}_{+}^{T}
\right) / 2 = \left(\mathcal{M}_{-}^{T} + \mathcal{M}_{-}
\right) / 2$, which is clearly symmetric (and real).

In turn, the lattice LCG condition can be obtained by
minimizing the functional \cite{LCGlattice}
\begin{equation}
\mathcal{E}_{LCG}[U_{\mu}, g, \Lambda] \, = \,
\mathcal{E}_{LG}[U_{\mu},g] \, + \, \Re \,\Tr \sum_x
     \,  i\, g(x) \, \Lambda(x) \; ,
\label{eq:LCGfunct}
\end{equation}
where the functions $\Lambda^b(x)$ are real-valued, generated
using a Gaussian distribution with width $\xi^{1/2}$.
By considering a one-parameter subgroup
\begin{equation}
g(x,\tau) \, = \, \exp \left[ i \tau
\sum_{b=1}^{N_c^2-1}\, \gamma^{\,\,b}(x) \lambda^{b} \right] \; ,
\label{eq:g}
\end{equation}
it is easy to check that the stationarity condition implies the lattice
LCG condition
\begin{equation}
 \nabla \cdot A^{b}(x)
\, \equiv \, \sum_{\mu=1}^{d} \, A_{\mu}^{b}(x+e_{\mu}/2) 
         \, - \, A_{\mu}^{b}(x-e_{\mu}/2) \, = \,
 \Lambda^{b}(x) \; ,
\label{eq:LCGcondition}
\end{equation}
which gives, in the formal continuum\footnote{As usual, we
indicate with $a$ the lattice spacing.} limit $a \to 0$,
the gauge condition $a\,\sum_{\mu} \partial_\mu A_\mu^b(x)
= \Lambda^b(x)$.
It is important to stress that the LCG functional
$\mathcal{E}_{LCG}[U_{\mu}, g, \Lambda]$ is linear in the
gauge transformation $\{ g(x) \}$.
This allows us to naturally extend to the LCG case the gauge-fixing 
algorithms used in Landau gauge \cite{CSD}.
On the other hand, the standard compact discretization (\ref{eq:defAgen})
for the gauge field $A^b_{\mu}(x+e_{\mu}/2)$ implies that this field is
bounded, so that $\nabla \cdot A^{b}(x)$ is also a bounded quantity.
This is in contrast with the $\Lambda^{b}(x)$ functions, which
are generated using a Gaussian distribution and, therefore,
are unbounded.
Thus, one can face convergence problems when a numerical
implementation of the LCG is attempted \cite{LCGlattice,
Cucchieri:2011aa,LCGgluon2}.

The continuum gauge field $\hat{A}_{\mu}(\vec{x})$ is usually
defined through the relation
\begin{equation}
U_{\mu}(\vec{x}) \, \equiv \,
\exp{\left[ i\, a\, g_0\, \hat{A}_{\mu}(\vec{x}+\vec{e}_{\mu}/2)
\,\right]} \; ,
\end{equation}
where $g_0$ is the bare coupling constant.
Then, Eq.\ (\ref{eq:LCGcondition}) yields
$a^2 g_0 \sum_{\mu=1}^d \left[
\partial_{\mu} \hat{A}_{\mu}^{b}(\vec{x})
   + \mathcal{O}(a^2) \right] = \Lambda^b(\vec{x})$
in the formal continuum limit $ a \to 0 $.
On the other hand, the usual gauge field in the continuum limit
---i.e.\ when the generators $\tilde{\lambda}_b$ are
considered [see Eq.\ (\ref{eq:lambda})]---is given by
\begin{equation}
2\,\hat{A}^{b}_{\mu}(\vec{x})\, \approx\, 2\, A^{b}_{\mu}(\vec{x})
/ \left( a\,g_0 \right) \; .
\label{eq:Aconti}
\end{equation}
Thus, with our notation, the continuum functions $\hat{\Lambda}^b(\vec{x})$
satisfy the relation $ 2\sum_{\mu=1}^d \partial_{\mu}
\hat{A}_{\mu}^{b}(\vec{x}) = \hat{\Lambda}^b(\vec{x})$
and we obtain
\begin{equation}
a^2\, g_0\, \hat{\Lambda}^b(\vec{x}) \,\approx\, 2\,
 \Lambda^b(\vec{x}) \; .
\end{equation}
Then, in the limit $\,a \to 0$, it is easy to show
\cite{Cucchieri:2018doy} that the expression
\begin{equation}
\frac{1}{2\xi} \sum_{x} \sum_{b=1}^{N_c^2-1}
\left[ \, \Lambda^b(x) \, \right]^2
\,\equiv\, \frac{\beta}{N_c\,\hat{\xi}} \sum_{x} \sum_{b=1}^{N_c^2-1}
\frac{\left[ \, 2\,\Lambda^b(x) \, \right]^2}{4}
\end{equation}
becomes
\begin{equation}
\frac{1}{2\,\hat{\xi}} \, \int d^dx \, \sum_{b=1}^{N_c^2-1} \,
[ \hat{\Lambda}^b(x) ]^2 \; .
\end{equation}
Here, $\beta = 2 N_c / (a^{4-d} g_0^2)$ in the lattice parameter
entering the Wilson action for the SU($N_c$) case, in a generic
$d$-dimensional space.
Thus, the continuum and lattice widths, $\hat{\xi}^{1/2}$ and
$\xi^{1/2}$, of the corresponding Gaussian distributions are
related through the expression $ \xi\equiv\hat{\xi}\,
N_c/(2 \beta)$.
This gives $ \xi \,<\, \hat{\xi}$ when $N_c \,< \, 2\,\beta$.
This inequality is satisfied for $N_c = 2,3$ and for typical
values of $\beta$ in the scaling region.

A few numerical studies of the gluon propagator have been
carried out, using the above lattice formulation for LCG
\cite{LCGlattice,LCGgluon2}.
In particular, it has been checked, for the SU(2) and SU(3) gauge
groups, that the longitudinal propagator $D_l(p^2)$ satisfies the
relation $p^2 D_l(p^2) = \xi$, as predicted by perturbation
theory.
At the same time, the transverse gluon propagator $D_t(p^2)$
has shown a clear dependence on the gauge parameter $\xi$, i.e.\
$D_t(0)$ decreases as $\xi$ increases.
Finally, $D_t(0)$ decreases if the lattice volume $V$ increases
\cite{LCGlattice}, as in Landau gauge.
These results are in agreement with the numerical data obtained
in Refs.\ \cite{Giusti:1999im,LCGgluon3}, using a different
formulation for the lattice LCG, and with several analytic
predictions \cite{Capri:2015pja,gluon,Siringo:2018uho}.

%%%%%%%%%%%%%%%%%%%%%%%%%%%%%%%%%%%%%%%%%%%%%%%%%%%%%%%%%%%%%%%%%%%%%%%%%%%%%%%%%%%%%

\section{The Ghost Sector}
\label{eq:ghost}

In order to define the ghost sector in lattice minimal LCG,
one should first recall that, in the continuum, there are in
principle three different possible setups for the LCG
(see, e.g., Appendix A in Ref.\ \cite{Alkofer:2000wg}):
\begin{enumerate}
\item[1)] complex ghost fields $\overline{c} = c^{\dagger}$, giving
the the FP matrix $- \partial \cdot D^{bc} $ and a non-Hermitian
Lagrangian density;
\item[2)] complex ghost fields $\overline{c} = c^{\dagger}$
and a symmetric FP matrix $- (\partial \cdot D^{bc} + D^{bc} \cdot
\partial) / 2 $, with a quartic ghost self-interaction term
in the Lagrangian density;
\item[3)] real independent ghost/anti-ghost fields $u, iv$ and
the effective Hermitian FP matrix
\begin{equation}
\frac{i}{2} \, \left( \begin{array}{cc} 0
          & - \, \partial \cdot D^{bc} \\[2mm]
   D^{bc} \cdot \partial & 0  \end{array} \right)
\,\equiv\, i\, M \; .
\label{eq:DD}
\end{equation}
\end{enumerate}

On the lattice, if one follows the same procedure used in Landau
gauge ---i.e.\ if one evaluates the second variation of the functional
${\cal E}_{LCG}[U_{\mu}, g, \Lambda]$, defined in Eq.\ (\ref{eq:LCGfunct}),
using the one-parameter subgroup (\ref{eq:g})---it is immediate to
see \cite{Cucchieri:2018doy,Cucchieri:2018leo,thesis} that the
term $\,i\, g(x) \, \Lambda(x)\,$ does not contribute to the FP matrix.
Thus, one is left with the second variation of the (Landau-gauge) term
${\cal E}_{LG}[U_{\mu},g]$, yielding the usual symmetric Landau FP matrix
$\mathcal{M}$, defined in Eq.\ (\ref{eq:MFP}).
As we have seen in the previous section, this FP matrix can also be
written as
\begin{equation}
\mathcal{M} \, = \, - \, \sum_{\mu=1}^d
 \frac{1}{2}\,\left[\,
    \nabla^{(-)}_{\mu} \,
        D_{\mu} \, + \, D^{T}_{\mu} \,
   \left(\nabla_{\mu}^{(-)}\right)^{T} \, \right]
        \; ,
\end{equation}
which corresponds to the symmetric FP matrix of case 2) above.
On the other hand, it is not clear how a quartic ghost
self-interaction term could be obtained on the lattice using
the approach considered here.
Let us stress that the above matrix $\mathcal{M}$ has real
non-negative eigenvalues and real eigenvectors, since it is
real and symmetric (and, therefore, it is Hermitian).

A possible lattice discretization of the FP matrix
$- \partial \cdot D^{bc} $ of case 1) is given by the
matrix in Eq.\ (\ref{eq:M+}), which can be written as
\begin{equation}
\mathcal{M}_{+}^{bc}(x,y) \,\equiv\,
\mathcal{M}^{bc}(x,y)\, +\, \sum_{e=1}^{N_c^2 -1}
\, f^{bec} \Lambda^{e}(x) \delta_{x,\,y} \; ,
\label{eq:mp}
\end{equation}
where we used the gauge condition (\ref{eq:LCGcondition})
and $f^{bce}$ are the (real) structure constants of the
SU($N_c$) gauge group, defined through the commutation
relations $ \left[\lambda^b, \lambda^c \right] \equiv
 2 i \sum_{e=1}^{N_c^2 -1} f^{bce} \lambda^e$.
On the other hand, the extra term in Eq.\ (\ref{eq:mp}) is
skew-symmetric under the simultaneous exchanges $b \leftrightarrow
c$ and $x \leftrightarrow y$, and it cannot be obtained
from a second variation, i.e.\ it should be added by hand!
It is important to note that the matrix $\mathcal{M}_{+}$
has complex-conjugate eigenvalues (and eigenvectors) with
a non-negative real part.
See Ref.\ \cite{Cucchieri:2018doy,Cucchieri:2018leo} for more
details about this setup.

Finally, when considering the continuum case 3), one should
note that the effective FP matrix $M$, defined in Eq.\
(\ref{eq:DD}), is also real and skew-symmetric and, therefore,
it cannot be obtained directly from a second variation of
any minimizing functional.
On the other hand, since $\mathcal{M}^{bc}(x,y)$ is real
and symmetric, if one extends to the complex case
the bilinear form
\begin{equation}
\sum{x,y} \sum_{b,c=1}^{N_c^2-1} \gamma^{\,\,b}_1(x)
\mathcal{M}^{bc}(x,y) \gamma^{\,\,c}_2(y) \; ,
\end{equation}
i.e.\ if one considers $\gamma_1^{\,\,b}(x), \gamma_2^{\,\,b}(x)
\in \mathbb{C}$, then the corresponding sesquilinear form
\begin{equation}
\left( \begin{array}{cc} \mathcal{M}\, & i\,\mathcal{M} \\[2mm]
   -i\, \mathcal{M} & \,\mathcal{M}  \end{array} \right)
\end{equation}
is a positive semi-definite Hermitian form.
Moreover, its imaginary part is skew-symmetric and gives
us a natural way of obtaining the FP matrix
\begin{equation}
M\,=\,\frac{1}{2} \, \left( \begin{array}{cc} 0
          & \mathcal{M}_{+} \\[2mm]
   - \mathcal{M}_{+}^T & 0  \end{array} \right) \,=\,
\frac{1}{2} \, \left( \begin{array}{cc} 0
          & \mathcal{M}_{+} \\[2mm]
   - \mathcal{M}_{-} & 0  \end{array} \right) \; ,
\label{eq:DD2}
\end{equation}
which is a possible discretization of the matrix
defined in Eq.\ (\ref{eq:DD}).
In this case, since $M$ is skew-symmetric, its eigenvalues
are complex-conjugate and purely imaginary, and they are
related to the singular-value decomposition of
$\mathcal{M}_{+}$, i.e.\ to the eigenvalues of
$\mathcal{M}_{+}^T \mathcal{M}_{+}$.

%%%%%%%%%%%%%%%%%%%%%%%%%%%%%%%%%%%%%%%%%%%%%%%%%%%%%%%%%%%%%%%%%%%%%%%%%%%%%%%%%%%%%

\section{Numerical Simulations: Ghost Propagator}

We have done some preliminary tests, evaluating the ghost
propagator using the FP matrix defined in Eq.\ (\ref{eq:mp}).
Since the matrix $\mathcal{M}_{+}^{bc}(x,y)$ is real and
not symmetric, we cannot use the conjugate gradient algorithm,
as in Landau gauge.
Thus, the inversion of the above FP matrix has been done
using \cite{Saad} the bi-conjugate gradient stabilized
algorithm, in the SU(2) case, and the generalized conjugate
residual, in the SU(3) case.
In both cases we used a point source \cite{source}.
Simulations have been carried out with $\beta = 2.4469$ for SU(2)
and $\beta = 6.0$ for SU(3), both corresponding
\cite{Cucchieri:2007zm} to a lattice spacing $a \approx 0.1\;
\mbox{fm}$.
Preliminary results for these two gauge groups have been
presented in Ref.\ \cite{Cucchieri:2018doy,Cucchieri:2018leo}.
One clearly sees from these data that the ghost propagator
in LCG agrees, within error bars, with the ghost propagator
in Landau gauge (for the same lattice setup).
This is in qualitative agreement with the theoretical
predictions of Ref.\ \cite{Siringo:2014lva} but in disagreement
with the finding of Refs.\ \cite{gluon,ghost}.

%%%%%%%%%%%%%%%%%%%%%%%%%%%%%%%%%%%%%%%%%%%%%%%%%%%%%%%%%%%%%%%%%%%%%%%%%%%%%%%%%%%%%

\section{Conclusions}

The numerical evaluation of the ghost propagator in LCG, using
the FP matrix $\mathcal{M}_{+}^{bc}(x,y)$ defined in Eq.\
(\ref{eq:mp}), seems feasible.
For the lattice setup considered in the simulations presented
in Ref.\ \cite{Cucchieri:2018doy,Cucchieri:2018leo}, the results are
essentially in agreement with the corresponding data in Landau
gauge.
Of course, simulations at larger physical volumes and different
gauge-fixing parameters $\xi$ should be done before one can conclude
that this is indeed the case.
At the same time, it would be important to extend these numerical
simulations to the cases 2) and 3), discussed in Sec.\ \ref{eq:ghost}.
Finally, one should try to understand how the first
Gribov region $\Omega$ can be defined in lattice minimal LCG,
i.e.\ if the GZ approach can be extended to the LCG on the lattice.

%%%%%%%%%%%%%%%%%%%%%%%%%%%%%%%%%%%%%%%%%%%%%%%%%%%%%%%%%%%%%%%%%%%%%%%%%%%%%%%%%%%%%

\acknowledgments

A.C.~and T.~M.~acknowledge partial support from CNPq.
A.C.\ also acknowledges partial support from FAPESP (grant
$\#$ 16/22732-1).
The research of D.D.~and M.R.~is supported by KU
Leuven IF project C14/16/067.
O.O.~and P.J.S.~acknowledge the Laboratory for
Advanced Computing at University of Coimbra
({\tt http://www.uc.pt/lca}) for providing access to the
HPC computing resource {\tt Navigator}.
P.J.S.\ acknowledges support by FCT under contracts
SFRH/BPD/40998/2007 and SFRH/BPD/109971/2015.
The SU(3) simulations were done using the {\tt Chroma}
\cite{Edwards:2004sx} and {\tt PFFT}~\cite{Pippig2013}
libraries.

%%%%%%%%%%%%%%%%%%%%%%%%%%%%%%%%%%%%%%%%%%%%%%%%%%%%%%%%%%%%%%%%%%%%%%%%%%%%%%%%%%%%%

\end{document}